# Interfacial charge trapping and chemical properties of deposited SiO$_2$ layers in 4H-SiC MOSFETs subjected to different nitridations

Patrick Fiorenza[1]*, Corrado Bongiorno[1], Filippo Giannazzo[1], Mario S. Alessandrino[2], Angelo Messina[1,2], Mario Saggio[2] and Fabrizio Roccaforte[1]

1) Consiglio Nazionale delle Ricerche – Istituto per la Microelettronica e Microsistemi (CNR-IMM), Strada VIII, n. 5 - Zona Industriale, 95121 Catania, Italy

2) STMicroelectronics, Stradale Primosole n. 50 - Zona Industriale, 95121 Catania, Italy

*E-mail: patrick.fiorenza@imm.cnr.it

## Abstract

In this paper, SiO$_2$ layers deposited on 4H-SiC and subjected to different post deposition annealing (PDA) in NO and N$_2$O were studied to identify the key factors influencing the channel mobility and threshold voltage stability in 4H-SiC MOSFETs. In particular, PDA in NO gave a higher channel mobility (55 cm$^2$V$^{-1}$s$^{-1}$) than PDA in N$_2$O (20 cm$^2$V$^{-1}$s$^{-1}$), and the subthreshold behavior of the devices confirmed a lower total amount of interface states for the NO case. This latter could be also deduced from the behavior of the capacitance-voltage characteristics of 4H-SiC MOSFETs measured in gate controlled diode configuration. On the other hand, cyclic gate bias stress measurements allowed to separate the contributions of interface states (N$_{it}$) both on the upper and bottom parts of the 4H-SiC band gap and near interface oxide traps (NIOTs) in the two oxides. In particular, it was found that NO annealing reduced the total density of charges trapped at the interface states down to 3 × 10$^{11}$ cm$^{-2}$ and those trapped inside the oxide down to 1 × 10$^{11}$ cm$^{-2}$. Electron energy loss spectroscopy demonstrated that the reduction of these traps in the NO annealed sample is due to the lower amounts of sub-stoichiometric silicon oxide (~ 1nm) and carbon-related defects (< 1nm) at the interface, respectively. This correlation represents a further step in the comprehension of the physics of the SiO$_2$/4H-SiC interface explaining the mobility and threshold voltage behavior of 4H-SiC MOSFETs.

INTRODUCTION

Silicon carbide (4H-SiC) is an excellent wide band gap semiconductor for the fabrication of metal oxide semiconductor field effect transistors (MOSFETs) suitable for highly efficient power switching applications [1]. However, the device behavior is strongly influenced by the processing of the $SiO_2$/4H-SiC interface and, consequently, several concerns still affect 4H-SiC MOSFET technology [2]. Among them, threshold voltage ($V_{th}$) instability phenomena [3,4] and channel mobility issues [5,6] often occur in 4H-SiC MOSFETs and are object of intensive debate. These effects can be correlated to charge trapping into and out of near-interfacial oxide traps (NIOTs) spatially located inside the gate oxide [7,8] or trapping at the interface states that reduce the free carriers density in the device inversion channel [9].

Due to the fact that silicon oxide ($SiO_2$) is the product of thermal oxidation of SiC, a great part of the literature works is focused on 4H-SiC MOSFETs employing thermal oxides ($SiO_2$) grown under different conditions. Typically, these oxides are subjected to post-oxidation annealings (POAs) to mitigate the presence of residual carbon at the interface. In fact, carbon clusters [10] and/or a carbon-rich transition layer at the $SiO_2$/SiC interface formed during thermal oxidation [11], oxygen vacancies [12] or nitrogen-related defects [13] are the main features commonly observed in these gate oxides. On the other hand, the interface between a *deposited* $SiO_2$ layer and 4H-SiC is, as such, not affected by the presence of carbon-related defects. However, the post deposition annealing (PDA) treatments typically needed to passivate the interface states [14] may induce a re-oxidation of 4H-SiC, accompanied by a certain intermixing at the $SiO_2$/4H-SiC interface.

A large variety of nitridations (i.e. in NO or $N_2O$) are typically employed to optimize the $SiO_2$/4H-SiC interface electrical properties [5], where the interfacial nitridation kinetics results from a balance between N incorporation and removal. In general, annealing in $N_2O$ improves the properties of the $SiO_2$/4H-SiC interface, but lesser than annealing in NO [15]. This is because $N_2O$ decomposes at high temperature into NO, $O_2$, and $N_2$. Hence, while the resulting NO produces a beneficial effect at the 4H-SiC/$SiO_2$ interface, the background oxygen can induce a re-oxidation, generating the undesired carbon-related defects at the interface. However, there is no full consensus in literature on which nitridation allows to obtain the best channel mobility in 4H-SiC MOSFETs. In fact, while typical channel mobility values of about 40 cm$^2$V$^{-1}$s$^{-1}$ are obtained after annealing in NO [7,9], some work reports excellent results also after N$_2$O annealing, with mobility values up to 99 cm$^2$V$^{-1}$s$^{-1}$ [16]. These contradictory results demonstrate that the intimate SiO$_2$/4H-SiC interface nitridation mechanisms are not fully understood yet.

In this context, it is still unclear which are the key physical factors influencing the final 4H-SiC MOSFET behavior, in terms of channel mobility with threshold voltage V$_{th}$ stability, respectively. Hence, correlating the chemical order of the *deposited* SiO$_2$/4H-SiC interfaces subjected to PDA and the 4H-SiC MOSFETs behavior is a key topic, which deserves further attention. For this purpose, while the characterization of simple MOS capacitors is unable to probe the complex energy levels' distribution inside the semiconductor band gap [17], appropriate measurement procedures applied to 4H-SiC MOSFETs can be useful to distinguish the effects of interface states (N$_{it}$) and near interface oxide traps (NIOTs) [18,19,20].

In this paper, SiO$_2$/4H-SiC interfaces, formed with deposited SiO$_2$ subjected to different post deposition annealings (PDAs) in NO or N$_2$O, were investigated by cyclic gate stress measurements and nanoscale structural/chemical analyses, with the aim to identify the key factors influencing the channel mobility and V$_{th}$ stability in lateral MOSFETs.

EXPERIMENTAL

Lateral MOSFETs were fabricated on 4°-off-axis n-type (0001) 4H-SiC epitaxial layers (1×10$^{16}$ cm$^{-3}$) and an Al-implanted body region (N$_A$~10$^{17}$cm$^{-3}$). The gate oxide was a 40 nm thick deposited SiO$_2$ layer [21]. After the gate oxide deposition at 700 °C, one sample was subjected to PDA in NO and the other in N$_2$O both at 1150 °C [22]. Due to the different reactivity of NO and N$_2$O, the processes durations were tuned in order to obtain a comparable amount of nitrogen at the interface

(as verified by secondary ion mass spectrometry (SIMS) [22]) and, hence, similar counter doping effect beneficial for the mobility [23,24]. The current voltage ($I_D$-$V_G$) transfer characteristics and capacitance-voltage (C-V) curves of the devices were measured in a CASCADE Microtech probe station, using a Keysight B1505A parameter analyser.

Scanning transmission electron microscopy (STEM) analyses were performed in a state of the art (Cs)-probe corrected JEOL ARM200CF at a primary beam energy of 200 keV operating in scanning mode. Electron energy loss spectroscopy (EELS) images are collected with a Gatan Quantum spectrometer in dual EELS configuration for energy drift correction. The energy dispersion was set to 0.25 eV/pixel in order to have all the three elements edge (100 eV for silicon, 285 eV for carbon and 530 eV for oxygen) in the same spectrum. TEM samples were prepared by mechanical polishing followed by low energy (0.5 keV) ion milling. Finally, the lamella was treated for 4 minutes in plasma $O_2$ to remove the residual surface carbon contamination.

RESULTS AND DISCUSSION

Fig. 1a shows the transfer characteristics ($I_D$-$V_G$) of the lateral 4H-SiC MOSFETs at a fixed drain voltage ($V_{DS}$ = +0.1V) by increasing the gate bias ($V_G$) above the threshold voltage ($V_{th}$). As can be seen, the sample annealed in NO shows a larger drain current compared to that annealed in $N_2O$ for the same $V_G$-$V_{th}$.

Fig. 1b reports the field effect channel mobility $\mu_{FE}$ determined on the two samples, defined as:

$$\mu_{FE} = \frac{L}{W C_{ox} V_{DS}} \frac{\partial I_D}{\partial V_G} \qquad (1)$$

where $W$ is the channel width, $V_{DS}$ is the source-drain bias, $C_{ox}$ is the gate oxide capacitance and $\partial I_D/\partial V_{GS}$ is the MOSFET transconductance [7,25].

The field effect mobility $\mu_{FE}$ contains the physical information on the modulation of the channel conductivity by the application of the gate bias. While $\mu_{FE}$ does not give directly information on the free electrons contributing to the conduction in the channel, this parameter can be used to compare the effectiveness of the nitridation process in minimizing the amount of trapped charge.

From Fig. 1b, it can be seen that the maximum mobility value after annealing in NO (55 cm$^2$V$^{-1}$s$^{-1}$) was more than twice larger than the value measured in the sample annealed in N$_2$O (20 cm$^2$V$^{-1}$s$^{-1}$). Furthermore, from the I$_D$-V$_G$ transfer characteristics in Fig. 1(a), the subthreshold swing at V$_{DS}$= + 0.1 V is decreased from 600 mV/decade for the N$_2$O down to 280 and mV/decade for the NO annealed sample, thus indicating that the total density of interface states in the upper part of the 4H-SiC band gap is strongly reduced by the PDA in NO.

To get further insights on the SiO$_2$/4H-SiC interfaces properties, C-V characteristics were collected on 4H-SiC MOSFETs in the gate controlled diode configuration [18]. Fig. 2 shows the C-V curves collected on both the NO or N$_2$O annealed devices. As can be seen, the C-V curves show two branches: one in the positive gate bias region (where the device approaches the inversion and the Fermi level moves close to the 4H-SiC conduction band edge), and another in the negative gate bias region (where the system goes in accumulation and the Fermi level is close to the 4H-SiC valence band edge). Notably, both branches of the C-V curves of the NO annealed sample are steeper than those of the N$_2$O sample, thus indicating that the PDA in NO reduced the interface traps both close to the conduction and to the valence band edge of 4H-SiC.

To quantify this effect, we probed the trapping effect in the upper and bottom parts of the band gap by monitoring the V$_{th}$ variation upon cyclic variable gate bias stress.

In particular, the variation of V$_{th}$ after stress was determined by a single gate bias point measurement (V$_{G-read}$), using the procedure described in Ref. [20]. Fig. 3a and 3b show the variation of V$_{th}$ in the N$_2$O and NO samples, respectively. The dashed arrows indicate the direction of the applied bias stress in the cycle. In particular, the gate bias stress was first ramped from V$_G$ = 0 V up to + 30 V with steps of 5 V. At each step, the I$_D$ was measured at V$_{G-read}$ = + 8V, and the variation of V$_{th}$ was determined. Here, an increase of the positive gate bias stress up to +30V produces a gradual increase of the $V_{th}$, thus indicating that the electron trapping occurring close to the 4H-SiC conduction band does not reach the saturation. In the second part of the cycle, the gate bias stress was ramped from 30 V down to - 25 V and finally back to zero. While in the positive gate bias region (30 V – 0V) the $V_{th}$ remains almost constant, it rapidly drops for negative $V_G$ values until reaching a saturation. It is possible to notice that the magnitude of the $V_{th}$ variation ($\Delta V_{th}$) recorded from negative to positive gate bias values is about 1 V for the $N_2O$ sample and 0.5 V for the NO sample. The variations of the threshold voltage ($\Delta V_{th}$), measured between $V_G = -25$ V and $V_G = +30$ V, are related to the density of the total charge trapped at the interface states ($N_{it}$) by the equation:

$$N_{it} = \frac{\kappa \varepsilon_0 \Delta V_{th}|_{V_G=-25}^{V_G=+30}}{q t_{ox}} \qquad (2)$$

where $q$ is the electron charge, $t_{ox}$ and $\kappa$ are the $SiO_2$ thickness and relative permittivity, respectively, and $\varepsilon_0$ is the vacuum permittivity.

The total $\Delta V_{th}$ variation observed in the gate bias range – 25 V / + 30 V in the two samples correspond to a total trapped charge density of $6 \times 10^{11}$ cm$^{-2}$ and $3 \times 10^{11}$ cm$^{-2}$, for the $N_2O$ and NO sample, respectively. However, during the bias stress cycle, starting from the equilibrium condition ($V_G$=0 V), the Fermi level moves towards the 4H-SiC conduction band for positive gate bias and towards the 4H-SiC valence band for negative gate bias. Hence, it is possible to distinguish between the charges trapped at interface states close to the conduction or to the valence band of 4H-SiC. In particular, from the negative $\Delta V_{th}$ a much lower density of charge trapped close to the 4H-SiC valence band can be estimated for the NO sample ($2 \times 10^{11}$ cm$^{-2}$) with respect to the $N_2O$ sample ($4.8 \times 10^{11}$ cm$^{-2}$). From the positive $\Delta V_{th}$ the estimated interface trapped charge close to the 4H-SiC conduction band is and $1 \times 10^{11}$ cm$^{-2}$ for the NO sample and $1.2 \times 10^{11}$ cm$^{-2}$ for the $N_2O$ sample.

On the other hand, closing the cycling stress measurement at $V_G$= 0 V resulted into a final $V_{th}$ value, which is different from the original one, thus being an effect of the variation in the equilibrium charge density trapped in the NIOTs [20] given by:

$$NIOTs = \frac{\kappa \varepsilon_0 \Delta V_{th}|_{V_G=0}}{q t_{ox}} \qquad (3)$$

In Fig. 3, the residual variations due to the near interface oxide traps, indicated with $\Delta V_{th}$(NIOTs), are highlighted by black and red arrows for the $N_2O$ and NO sample, respectively. Using Eq. 3, from the residual threshold voltage variation $\Delta V_{th}$(NIOTs) it was possible to extract the amount of NIOTs in both samples. Noteworthy, while for the $N_2O$ sample the measured NIOTs was $1.5 \times 10^{11}$ cm$^{-2}$, this value was reduced down to $1 \times 10^{11}$ cm$^{-2}$ in the NO annealed device.

In order to clarify the different electrical behavior of the two oxides in relation to the physical nature of the traps, a nanoscale investigation of the *deposited* $SiO_2$ layer before PDA (used as reference) and of the same $SiO_2$ after being subjected to PDA in $N_2O$ and NO has been performed by means of STEM-EELS.

Figs. 4 shows a typical dark field STEM image collected on a $SiO_2$/4H-SiC nitridated interface. Here, it is possible to notice the presence of several terraces composed of 4H-SiC crystalline basal planes 4° inclined (due to the off-cut angle) with respect of the $SiO_2$/4H-SiC interface. Furthermore, Figs. 5a and 5b show carbon, oxygen and $SiO_x$ EELS profiles collected on both $N_2O$ and NO samples, in a spectrum box orthogonal to the basal plane (indicated by the dashed line in Fig. 4). The EELS profiles of the deposited $SiO_2$ without PDA (reference) are also reported to understand the effect of the re-oxidation induced by the PDA. As can be seen, both samples show the presence of a sub-stoichiometric $SiO_x$ profile obtained using a 4 eV energy window between 99 and 103 eV. In fact, in this energy range only the silicon atoms that are not completely surrounded by oxygen can give a contribution to the EELS spectrum above the detection limit.

It is possible to notice the progressive change in the $SiO_x$ and oxygen profiles across the $SiO_2$/4H-SiC interface, due to the slight re-oxidation induced by the PDA. In fact, the thickness of the sub-stoichiometric $SiO_x$ layer was reduced from ~ 2 nm in the $N_2O$ sample down to about 1 nm in the NO sample. In general, oxygen-related defects, or the presence of a $SiO_x$ layer, have been correlated to the presence of interface traps [20,26]. Furthermore, it is worth noticing that the intensity of the $SiO_x$ profile in the NO sample decreases more rapidity compared to the N$_2$O sample. Hence, our result demonstrates that the smaller amount of interface traps in the sample annealed in NO can be associated to a thinner and steeper sub-stoichiometric SiO$_x$ layer with respect to the N$_2$O case. In particular, in a theoretical work Zhang et al. recently pointed out that Si interstitials in the sub-stoichiometric SiO$_x$ layer can also be associated to the near interface oxide traps [27]. However, a clear experimental evidence is still missing.

On the other hand, in both nitridated samples the carbon profiles show a decreasing tail within the oxide. The decreasing carbon tails extend from the SiC interface into the oxide for about 1.2 nm for the sample annealed in N$_2$O (Fig. 5a) and for about 0.9 nm for the sample annealed in NO (Fig. 5b). These carbon-related defects within the insulator could be associated to the residual NIOTs detected by the cyclic gate bias stress procedure.

The scenario deduced by cross correlating the electrical measurements with the chemical analysis of the nitridated SiO$_2$/4H-SiC interface is represented in a schematic band diagram in Fig. 6, showing the energetic distribution of interface states ($N_{it}$) and near interface oxide traps (NIOTs) in the system. In particular, Fig. 6 schematically shows also how these SiO$_x$ and carbon-related defects can be associated with the interface traps located close to the 4H-SiC valence band edge [28] rather than inside the insulating layer [27]. Hence, the lower carbon tail observed upon PDA in NO is consistent with lower amount of NIOTs and the better $V_{th}$ stability under cyclic stress compared to the N$_2$O sample and the lower amount of the interface traps close to the valence band (Fig. 6).

CONCLUSION

In conclusion, SiO$_2$ layers deposited on 4H-SiC and subjected to different PDA in NO and N$_2$O were studied to identify the key factors influencing the channel mobility and threshold voltage stability in 4H-SiC MOSFETs. The higher channel mobility (55 cm$^2$V$^{-1}$s$^{-1}$) measured after PDA in NO with respect to that (20 cm$^2$V$^{-1}$s$^{-1}$) measured in N$_2$O was due to a lower total density of interface states for the NO case. The contributions of interface states ($N_{it}$) and near interface oxide traps

(NIOTs) in the two oxides could be distinguished by means of cyclic gate bias stress measurements. In particular, NO annealing reduced the total density of charge trapped at the interface states down to $3 \times 10^{11}$ cm$^{-2}$ and the charges trapped inside the oxide to $1 \times 10^{11}$ cm$^{-2}$. EELS analyses enabled to correlate the reduction of these traps to the lower amounts of sub-stoichiometric silicon oxide and carbon-based defects at the interface, respectively. The results explained the mobility and threshold voltage behavior of 4H-SiC MOSFETs, indicating that an accurate control of the nitridation conditions of deposited oxides is required to minimize the interfacial re-oxidation.

This work was carried out in the framework of the ECSEL JU project REACTION (first and euRopEAn siC eigTh Inches pilOt liNe), grant agreement no. 783158.

The data that support the findings of this study are available from the corresponding author upon reasonable request.

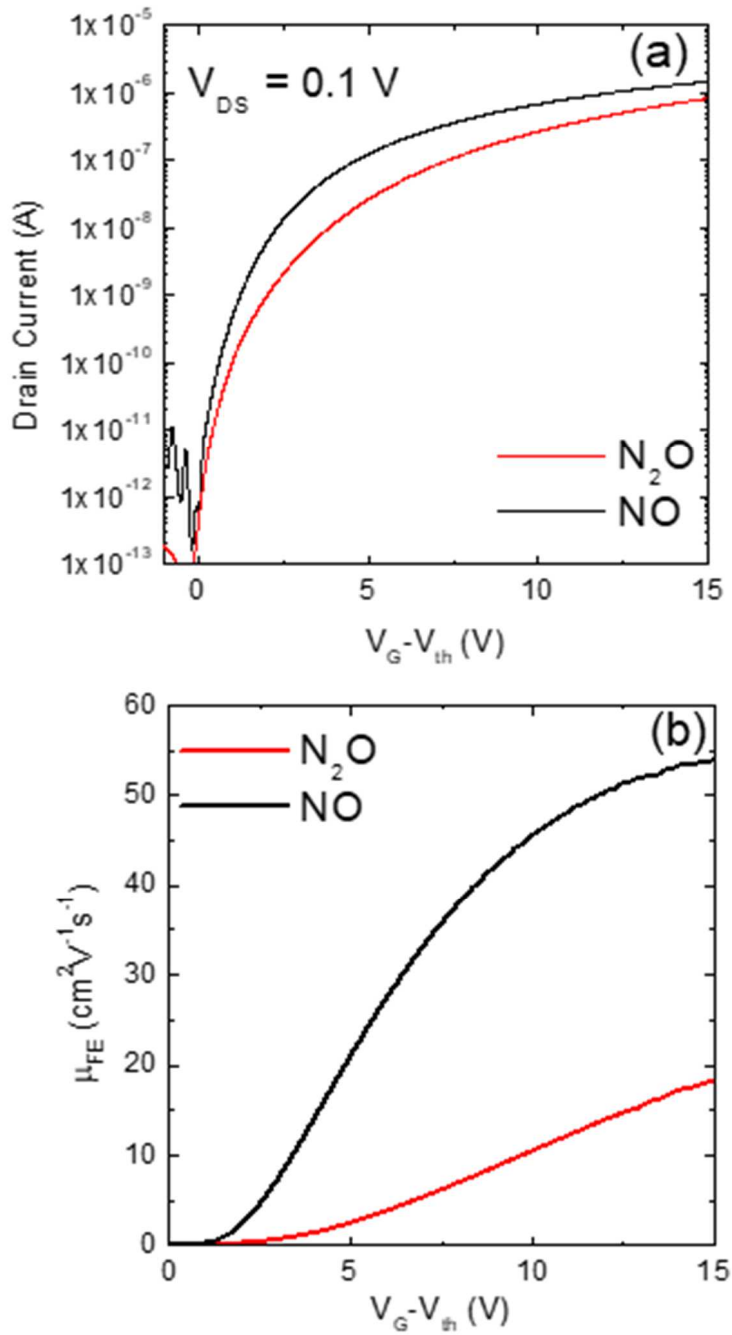

Fig. 1: (a) Drain current $I_D$ as a function of $V_G$-$V_{th}$ (at $V_{DS}$ = 0.1 V) for both NO and N$_2$O MOSFETs. (b) Field effect mobility $\mu_{FE}$ for both NO and N$_2$O MOSFETs.

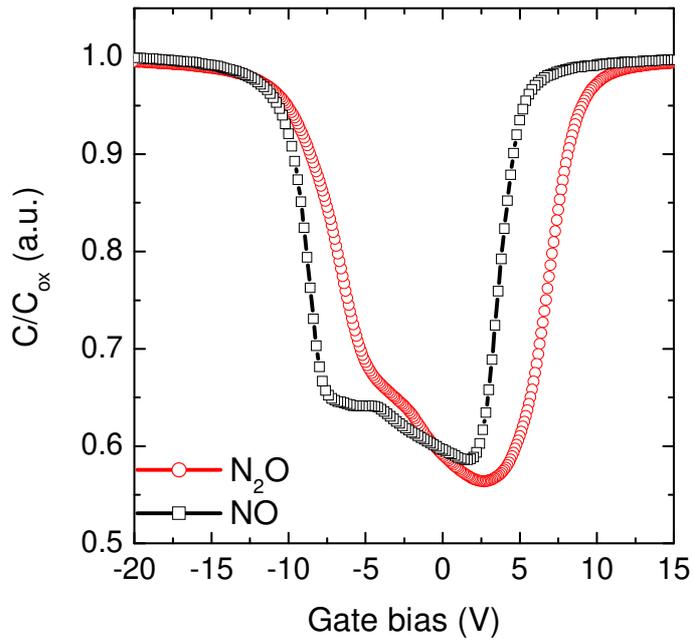

Fig. 2: C-V characteristics collected at 1 kHz on 4H-SiC MOSFETs in gate controlled diode configuration. The NO annealed device shows steeper branches compared to the $N_2O$, thus indicating a reduced amount of interface traps.

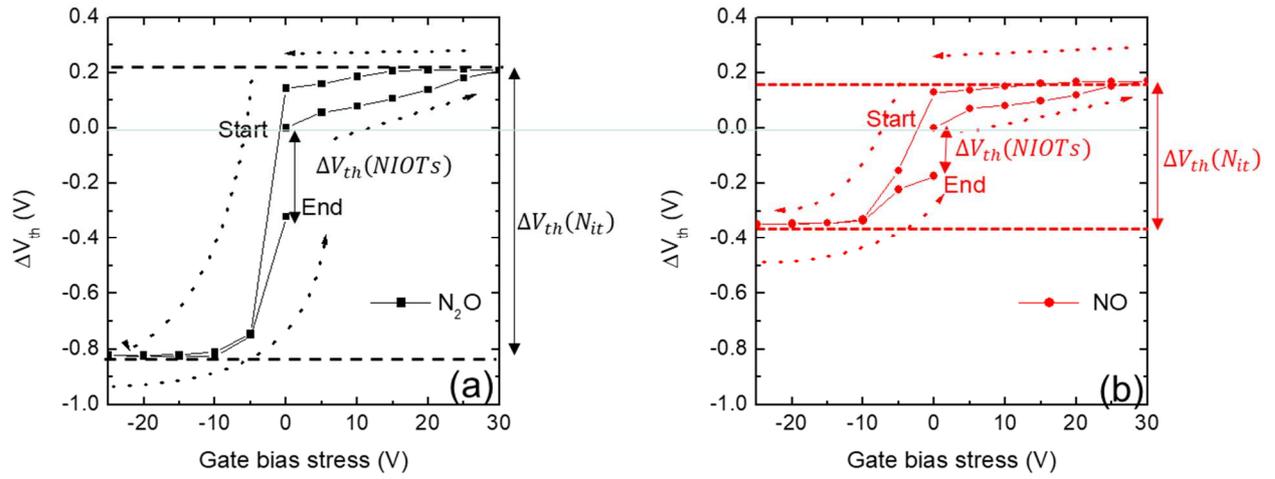

Fig. 3: Threshold voltage variation $\Delta V_{th}$ measured during the cyclic gate bias stress on $N_2O$ (a) and NO (b) samples, enabling the identification of the interface states ($N_{it}$) and near interface oxide traps (NIOTs) contributions. The dashed arrows indicate the direction of the applied bias stress in the cycle.

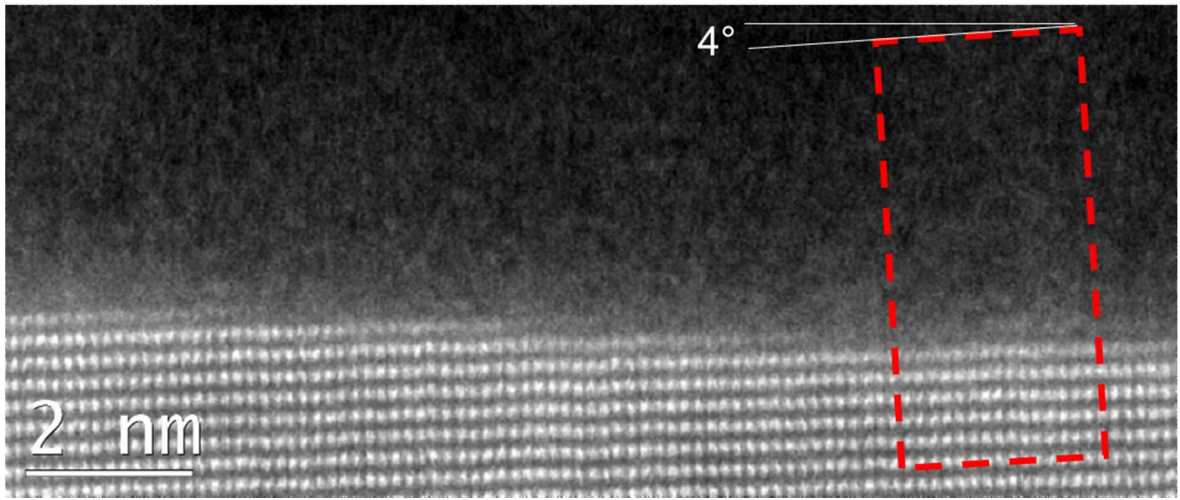

Fig. 4: High resolution STEM image across a typical SiO$_2$/4H-SiC nitridated interface. The EELS acquisition box is schematically indicated by the dashed line.

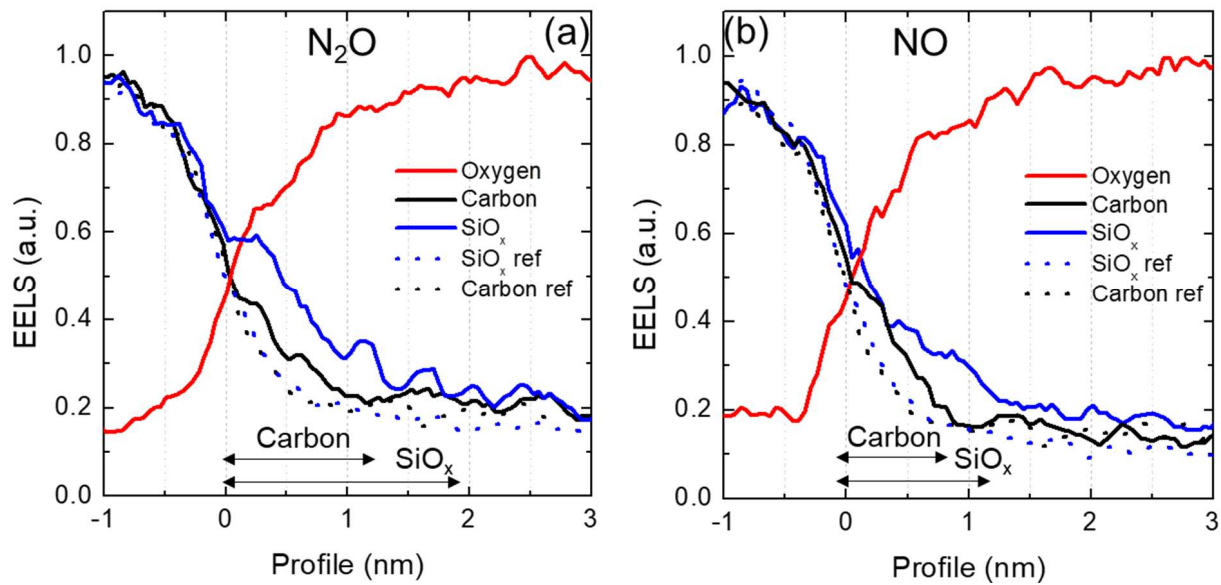

Fig. 5: EELS profiles of SiO$_x$, carbon and oxygen for the samples subjected to PDA in N$_2$O (b) and NO (c). The silicon and carbon profiles acquired on an as-deposited SiO$_2$ layer are reported as a reference.

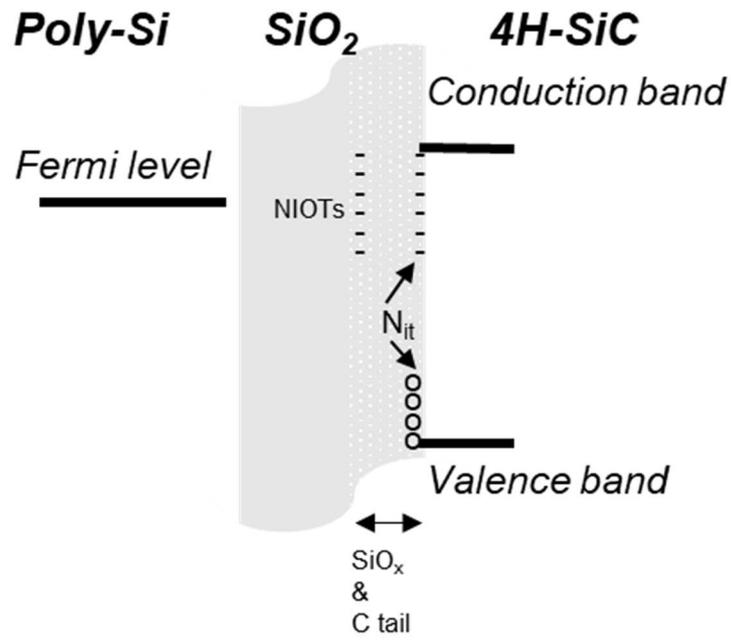

Fig. 6: Graphical representation of the $SiO_2$/4H-SiC interface, indicating the presence of interface states ($N_{it}$) close to the valence and conduction band edges, and near interface oxide traps (NIOTs) located in the upper part of the band gap. These traps are associated with $SiO_x$ and carbon-related defects.